\documentstyle[11pt,newpasp,twoside,epsf]{article}
\begin{document}

\title{Galactic Disk Warps}

\author{Konrad Kuijken and I\~nigo Garcia-Ruiz}
\affil{Kapteyn Institute, PO Box 800, 9700 AV Groningen,
the Netherlands}

\section{Introduction}

Warps of disk galaxies are a common phenomenon (as common as spirtal
structure), yet they are still not fully understood. In this review I
will try to summarize their observed properties, including preliminary
results of a new HI survey we have carried out with Westerbork, and
try to relate these to several proposed mechanisms for explaining
warps. I will not attempt a historic review of the theoretical work on
this subject; such reviews can be found in Binney (1992), or Kuijken
(1998), for example.

\section{Observed Properties}

The global observational understanding of galaxy warps has been
summarized in two pairs of Laws. 
\begin{itemize}
\item{\em Bosma's Laws} (Bosma 1991, on overall statistics of warps)
\begin{enumerate}
\item At least half of all galaxies are warped\\
       {\bf Implication:} Warps are long-lived or continuously
       generated
\item Galaxies with small dark halo core radii (as determined from a
       rotation curve decomposition) are less likely to be warped\\
	{\bf Implication:} Link between warps and the dark halo
	potential
\end{enumerate}
\item{\em Briggs's Laws} (Briggs 1990, on structure of individual warps)
\begin{enumerate}
\item Disks are generally flat inside radius $R_{25}$. Out to radius
	$R_{26.5}$ the line of nodes of a warp is straight\\
       {\bf Implication:} Self-gravity of the disk is important
       (it keeps the different parts of the disk precessing
       synchronously and hence the line of nodes straight---cf. the
       winding problem of spiral waves)
\item The outer line of nodes advances in the direction of galactic
       rotation\\
       {\bf Implication:} Warps are not quite in equilibrium at large
       radii. This points to a link to the environment, or to very
       long timescales
\end{enumerate}
\end{itemize}

Warps are primarily observed in HI. Significant warps (misalignments
between inner and outer parts of the disk of more than a degree or
two) in stellar disks are rarely observed (Reshetnikov \& Combes 1999), though
recently the warp of the Milky Way has also been seen in the stellar
distribution (Alard 2000).  . This may indicate that warps are a
phenomenon which affects only the cold ISM, or that only the very
outskirts of galaxy disks, which are only observable in HI, are
involved.

The Galaxy warp illustrates the general point that, analogously to
spiral structure, warps come in several varieties: `grand-design',
nice integral-sign shaped bi-symmetric warps; `irregular' warps which
are only visible on one side, or in which one side of the galaxy is
more warped than the other; and `feeble' (weak or absent) warps.  To
determine the relative frequency of these classes is important, as it
points the way for attempts at an explanation: should we be looking
for mechanisms which naturally produce beautiful integral-sign warps,
or for---possibily more chaotic---ways to make irregular ones?

In order to address this question, we have performed a blind survey of
edge-on (as judged from optical images) galaxies. These galaxies form
part of the WHISP sample, which is basically an HI flux-density
limited sample ($>200$mJy) selected from the UGC catalogue. We chose
galaxies with blue major diameters larger than 2', and inclination
class 6 or 7. DSS images of these galaxies were inspected to filter
out obviously less-inclined galaxies, to leave us with a sample of
galaxies which, as judged from optical images, have inclinations at
least $\sim80^\circ$. We chose edge-on galaxies for this survey so that
the warps can be studied purely morphologically, without the need for
interpretation or modelling of the velocity field (e.g., by means of
tilted ring models).

Our sample clearly shows that warps occur in all types. About 1/3 of 
our galaxies show a nice integral-sign warp, about 1/3
are flat, and the final 3rd of the sample are of the irregular
(one-sided or asymmetric) types (Table 1).

\begin{table}
\caption{The classification of warps in a sample of 28 edge-on
galaxies, observed in HI with Westerbork (Garcia-Ruiz et al., in
preparation). Note the large fraction of galaxies which do not fall in
the classical, integral-sign warp category.}

\begin{tabular}{lr}
\tableline
Warp type  & \multicolumn{1}{c}{Number} \\
\tableline
Flat & 7\\
Integral-sign & 9\\
Asymmetric & 8\\
U-shaped & 2\\
\tableline
Total & 26 \\
\tableline
\tableline
\end{tabular}

\end{table}

\section{Models}

Though there have been suggestions
that magnetic fields coupling to a slightly ionized ISM could explain
the observed warps (Battaner et al. 1991) we will concentrate here on
models invoking gravitational processes. The fact that regular disk
galaxies show no major discrepancies between stellar and gaseous
rotation curves (beyond the well-understood asymmetric
drift) indicates that magnetic fields are in any case not a dominant
contributor to large-scale galaxy dynamics. Moreover, in the Milky Way,
recently Alard (2000) has shown that the stellar disk shares in the
warp seen in the HI.

I will briefly discuss three scenarios for warps: the normal mode
models, accretion of angular momentum, and interaction with
satellites.

\subsection{Normal Modes}
The normal mode picture of warps (Toomre 1983; Dekel \& Shlosman 1983;
Sparke \& Casertano 1988) envisions the disks as embedded in a
flattened dark halo. If the disk is misaligned with the equator of the
halo, the resulting torque combined with the disk's spin will cause it
to precess about the halo minor axis. This precession rate is easily
calculated (Kuijken 1991) as the ratio of the torque from the halo
(determined from the circular and vertical frequencies $\Omega_H$ and
$\nu_H$ in the potential of the halo) to the spin of the disk
(governed by the total circular frequency $\Omega_c$):
\begin{equation}
\Omega_{prec}={\int \Sigma R^3 (\Omega_H^2-\nu_H^2) dR
              \over
	      {\int 2 \Sigma R^3 \Omega_c dR }}
\label{eq:sc}
\end{equation}
where $\Sigma$ is the surface density of the disk at radius $R$. 

Because different parts of the disk have a different free precession
frequency in the potential of the halo, without disk selfgravity the
warp would wind up and disappear, analogously to spiral perturbations in
massless disks. However, the disk does have selfgravity, and Sparke
and Casertano were able to show that solutions exist in which the disk
precesses as a single, warped, unit in the halo. When the precession
frequency difference across the disk is too large, or the
disk is too lightweight, these solutions do not exist. 

The problem with normal modes was already forseen by Toomre (1983),
who pointed out that a crucial assumption in this picture is the
neglect of backreaction from the halo to the precession of a massive
disk in its center. This effect, analogous to dynamical friction, was
investigated by Dubinski \& Kuijken (1995), Nelson \& Tremaine (1995),
and Binney et al. (1998) after earlier investigation in terms of WKB
density waves by Bertin and Mark (1980). It turns out to be very
important: the precession frequencies implied by equation \ref{eq:sc}
are so low compared to the dynamical time of the inner halo that there
is ample time for the inner halo to adjust itself to the reorientation
of the disk. This aligned inner halo results in a diminished torque,
which in turn further reduces the precession speed. The overall effect
is an alignment wave which propagates outwards through the disk, as
the precession grinds to a halt.  The alignment between disk and halo
is rather fast, on the order of three local orbital times.

\begin{figure}
\plotone{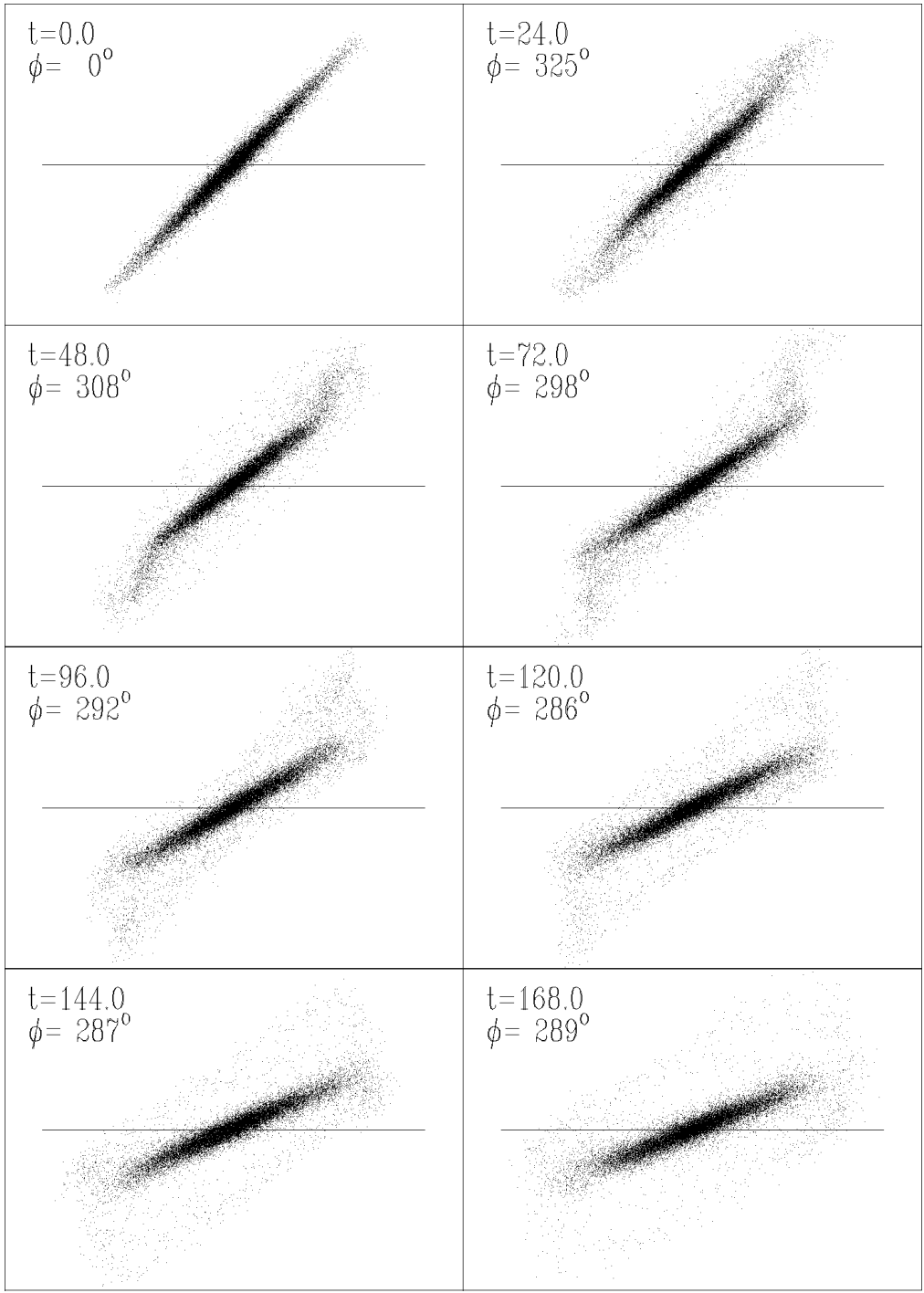}
\caption{$N$-body simulation of the evolution of an inclined disk in a
live dark halo (Dubinski \& Kuijken 1995). The azimuth from which the
disk is viewed is indicated in each panel. Note that the precession
very quickly comes to a halt, before even a quarter of a precession orbit
has been completed.}
\end{figure}

\begin{figure}
\plotfiddle{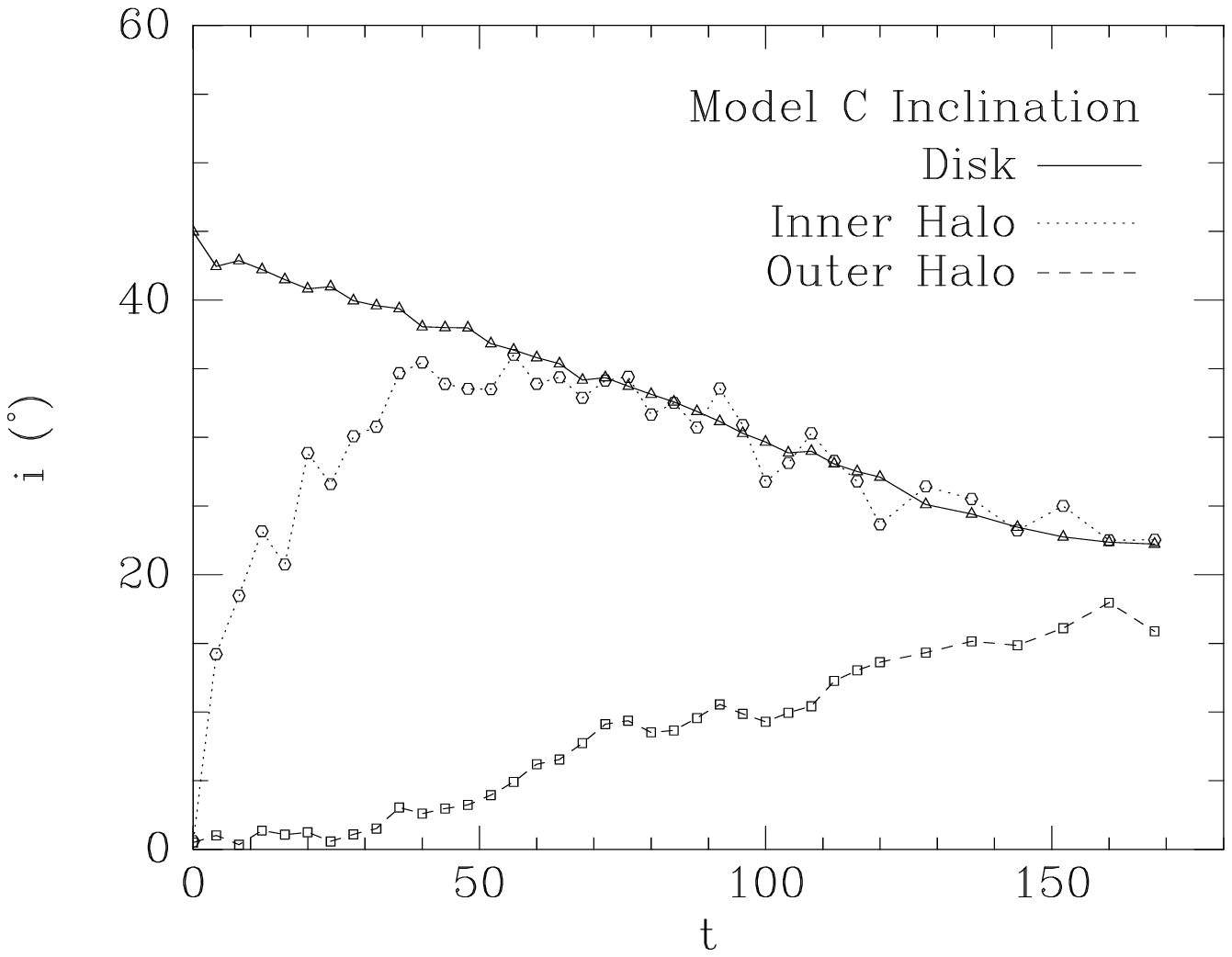}{4.5truecm}{0}{50}{50}{-142}{-127}
\caption{The evolution of the inclination of the
disk, the inner halo and the outer halo with time in the simulation
shown in fig 1. The inner halo
and the disk quickly align; over a longer time also the outer halo
inclination approaches that of the disk.}
\label{fig:dk}
\end{figure}

Under some conditions, the close coupling between halo and disk can
instead lead to a fast excitation. This can happen, for example, if
the halo rotates retrograde with respect to the disk, or if it is
prolate (Nelson \& Tremaine 1995).

\subsection{Accretion of Angular Momentum}

There is evidence that galaxies from time to time acquire material
with significantly different alignment of angular momentum. Polar
rings and counterrotating components in disks are clear
examples. 

If this such accretion continuously alters the orientation of the dark
halo angular momentum axis, the disk finds itself in a slowly changing
potential, in which it is continuously trying to align itself with the
dark halo symmetry plane. Damping is now not such a problem, as the
halo is continuously changing in response to outside influences,
rather than only to the disk. As shown by Jiang \& Binney (1999), this
process can generate realistic warp amplitudes in disks.

If this mechanism dominates, one might expect this slowly changing
tidal fields to result in rather symmetric warps, though no detailed
simulations have yet been performed.

\subsection{Forcing by Satellites}

One of the first possible explanations for the warp of the Galaxy was
the tidal perturbation by the Magellanic Clouds. Hunter \& Toomre
(1969) however showed that these are too weak to have a strong effect.
Lately this idea has been revived, however, by the work of Weinberg
(1995, 1998). He presented a model for which he calculated the
self-consistent response (wake) of the Galactic dark halo to the
orbiting Clouds. This wake turns out to be rather massive, and to be
strongest at a radius inside that of the Cloud orbit; both help to
boost the tidal effect of the wake to dominate that of the Cloud
itself. With such halo-enhanced tides, he was able to reproduce the
observed Galactic warp amplitude.

Weinberg's model requires some tuning in order to achieve the
amplitudes observed. It also makes several assumptions: 
\begin{enumerate}
\item The Cloud
orbit is modelled as a quasi-periodic orbit, i.e., the dynamical
friction-induced decay of the orbit is ignored. This means that the
model disk is
perturbed at a constant set of frequencies, and it has time to build
up its response to these perturbations. In reality, however, the frequencies at
which the disk is perturbed are continuously increasing.
\item The halo's back-reaction to the disk precession is not included
in the calculations. In fact, in these models the halo is assumed to be
spherically symmetric (before the LMC perturbs it). We have seen
above, however,
that the halo couples very effectively to a precessing misaligned disk.
\end{enumerate}
These caveats suggest that satellites may not
generally be responsible for causing warps in their hosts.

Some extra evidence against the LMC as cause comes from studying the
orientation of the warp and of the LMC orbit.  Garcia-Ruiz et al
(in prep.) made a very simple disk/halo/satellite system. The model
consists of a rigid spinning disk, a flattened dark halo, and a
satellite orbiting on a circular orbit. The equations of motion for
the polar coordinates $(\theta,\phi)$ of the disk axis (halo axis
$\theta=0$) can
then be linearized in terms of $(x,y)=(\theta\cos\phi,\theta\sin\phi)$
to 
\begin{eqnarray}
I_1 \ddot x + S \dot y + V_H x &=& V_S \sin2\theta_S(t)\cos\phi_S(t)\\
I_1 \ddot y - S \dot x + V_H y &=& V_S \sin2\theta_S(t)\sin\phi_S(t)
\end{eqnarray}
where $I_1$ is the moment of inertia of the disk about its diameter,
$S$ is the spin of the disk, $V_H$ is the torque per unit angle
exterted by the halo, $V_S$ parameterized the satellite tidal field
strength and the satellite orbit follows
$\theta=\theta_S(t),\phi=\phi_S(t)$. This equation is easily solved:
for the case of a polar orbit with $\phi_S=90^\circ$,
$\theta_S=\Omega_S t$ the solution is 
(with $\Delta=(V_H-4I_1\Omega_S^2)^2-(2\Omega_S S)^2$)
\begin{eqnarray}
x&=&-{2\Omega_S S\over\Delta} V_S\cos2\Omega_S t\\
y&=&{V_H-4I_1\Omega_S^2\over\Delta} V_S\sin2\Omega_S t
\end{eqnarray}

\begin{figure}
\plotfiddle{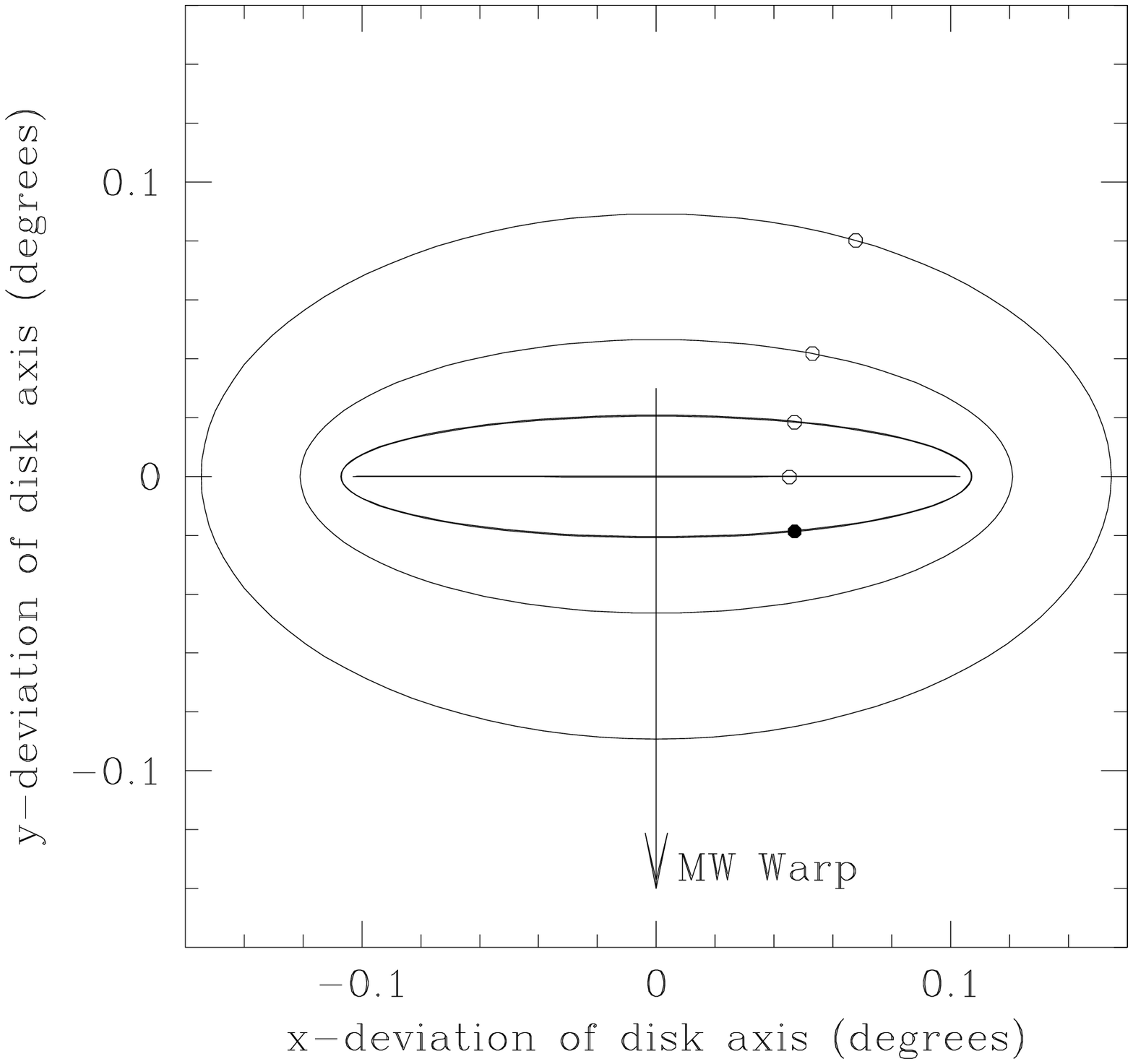}{4truecm}{0}{33}{33}{-200}{-88}
\plotfiddle{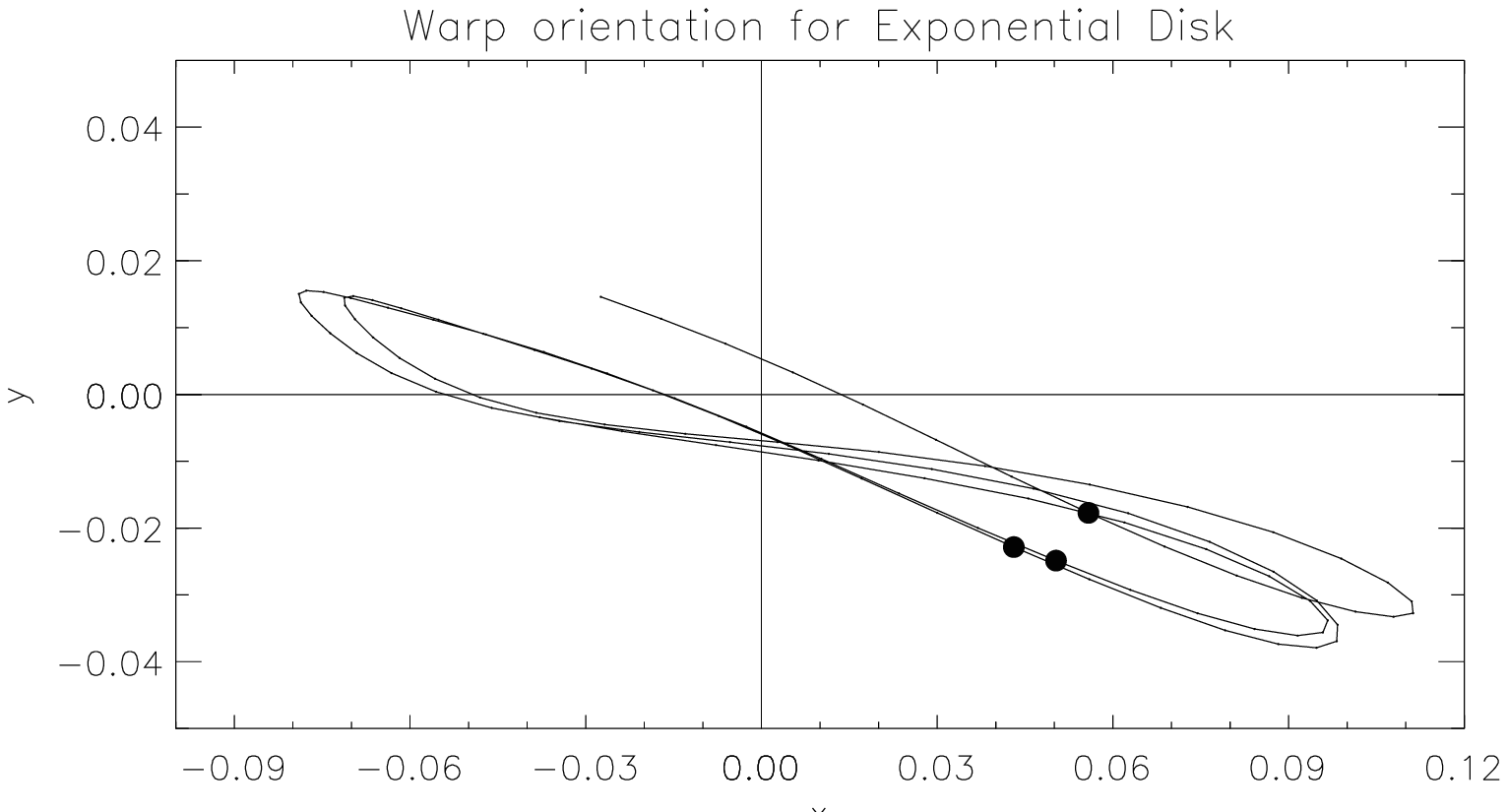}{0truecm}{0}{45}{45}{-15}{30}
\caption{Left: precession path of a rigid disk in a
flattened halo, perturbed by an orbiting satellite. The dots indicate
the position corresponding to the current phase of the LMC in its
orbit, which is in the $yz$ plane fro halo potential flattenings of 0
(solid), 0.05, 0.1, 0.15, 0.2.
The direction of the observed
tilt of the axis of the outer disk of the Milky Way is indicated by the
arrow. 
Right:
The warp direction (direction of the
average alignment of the outer disk with reference to the inner disk
plane) for an exponential disk consisting of centrally-pivoted
concentric spinning rings. In both cases, the response of the disk is
primarily to tilt in the plane perpendicular to the satellite orbit,
unlike what is observed in the Galaxy/LMC system.}
\label{fig:tilt}
\label{fig:tiltr}
\end{figure}

%

The precession path of such a rigid disk, for Galaxy/LMC parameters,
is shown in Fig~\ref{fig:tiltr}. The dominant effect of the satellite
perturbation is to make the disk axis nod perpendicular to the
satellite orbital plane, a consequence of the slow satellite orbital
frequency compared to the natural precession frequency of the
disk. The orientation is independent of the satellite mass, or of the
strength of the tidal field; it is governed only by the frequencies of
the satellite and disk. 

An extension of this calculation to a disk consisting of many rigid,
precessing rings all anchored to a common center shows the same trend
(Fig~\ref{fig:tilt}): the warp orientation (defined as the direction
along which the inner and outer disk orientations differ maximally) is
perpendicular to the satellite orbital plane.

By contract, the LMC orbit lies in the plane along which the Galactic
warp is strongest. 

A full N-body simulation of a disk/galaxy/LMC system, including halo
back-reaction on the disk, shows the same effects. It appears that the
wake of the satellite in the halo basically acts in phase with the
satellite, and so mostly changes the amplitude of the tides but not
their orientation. The backreaction of the halo on the precessing disk
is more difficult to predict in these analytical models, as it can act
both as a damping or as a destabilizing factor (see discussion on
normal modes above); our numerical experiments in simulating the
Galaxy/LMC situation have in not turned up any cases where the disk
response was as strong as observed in the Milky Way, or oriented as
observed.

Interestingly, the Sagittarius dwarf has an orbit almost at right
angles to that of the LMC. It could therefore in principle produce a
response in the disk of the right phase (but see Binney
2000). However, at this moment the mass of the Sgr dwarf is still
rather uncertain, and it is not clear whether it could raise
sufficiently strong tides. It is also in a sufficiently small orbit
that the tides will be very asymmetric (Ibata \& Razoumov 1998).

All in all, therefore, it appears that satellites as an explantion of
the warp of the Galaxy, and by extension as a generic model for warps
in other galaxies, pose some difficulties.

\section{Summary}

In spite of being known now for many decades, warps are still a
puzzle. The observational situation and theoretical models to explain
the data are still evolving. In this talk I have tried to summarize
the situation as follows.

\begin{enumerate}
\item Warps are a very common phenomenon. However, in addition to the
classic `grand-design' integral-sign warps, in many cases the warp is
asymmetric or even one-sided. Models for warps should therefore not
only address the symmetric, regular ones. Quite possibly there
is a good analogy here to spiral structure, which also displays varying
degrees of regularity.
\item Normal modes (equilibrium configurations of a tilted, warped
disk precessing about the symmetry axis of a flattened halo) interact
strongly with the dark halo. As a consequence they are easily and
strongly damped, or, in special circumstances, excited. 
\item Satellite tides are generally too weak to produce warps of the
amplitudes observed. However the dark halo can under certain
circumstances respond to the satellite in a way that 
significantly adds to the tidal perturbation on the disk. A useful
diagnostic for satellite tides as explanation for a warp is the
orientation of the warp with respect to the satellite orbit. In the
case of the LMC/Galaxy system, this orientation appears to be
different from predictions.
\item Accretion of material can generate warps through a continual
change in the orientation of the halo symmetry plane, to which
different parts of the disk respond on different timescales. Such
models are yet to be investigated in full detail; in particular the
question of whether the observed frequency of asymmetric warps can be
reproduced by this kind of model may be a useful avenue to explore.
\end{enumerate}

\end{document}